\def\compileforpublish{1}

\documentclass[letterpaper, 10pt, conference]{./IEEEtran}  

\IEEEoverridecommandlockouts                              

\usepackage{tabularx,calc}
\usepackage{color}
\usepackage{options}
\usepackage{hyperref}
\usepackage[pdftex]{graphicx}

\usepackage{tikz}
\usetikzlibrary{%
	 			 shapes,
	 			 automata,
	 			 arrows,
	 			 matrix,
	 			 backgrounds,
	 			 fit,
	 			 patterns,
	 			 decorations.markings, 
	 			 svg.path, 
	 			 shapes.multipart, 
	 			 external, 
	 			 shadows, 
	 			 fadings,
	 			 shadings,
	 			 positioning, 
	 			 calc, 
	 			 decorations, 
	 			 decorations.text,
	 			 intersections, 
	 			 angles, 
	 			 quotes,
   			 	 decorations.pathmorphing, 
   			 	 arrows.meta, 
   			 	 patterns, 
   			 	 decorations.pathreplacing, 
   			 	 calligraphy,
   			 	 external,
   			 	 snakes,
   			 	 bending,   			 	 
   			 	 babel%
}     		 	 
\usepackage{tkz-euclide}

\tikzexternaldisable 

\usepackage{pgfplots}
\usepackage{pgfplotstable}

\usepackage{tikz-3dplot}

\usepackage{siunitx}
\usepackage[
			bibstyle=ieee,
			citestyle=numeric-comp,
			backend=bibtex,
			minnames=1,
			maxcitenames=2,
			maxbibnames=99,
			doi=true,
			isbn=false,
			url=false,
			hyperref=true,
			natbib=true]
			{biblatex}

\DeclareDatamodelEntrytypes{standard}
\DeclareDatamodelEntryfields[standard]{type,number}
\DeclareBibliographyDriver{standard}{%
        \printfield{title}
	\printfield{shorttitle}\addcolon%
	\printfield{year}%
	\addspace
	\parentext{\printfield{year}}
	\setunit{\labelnamepunct}\newblock
	\usebibmacro{title}%
	\newunit\newblock
	\printnames{author}%
	\setunit{\addspace}\newblock
	\printfield[parens]{type}%
	\newunit\newblock
	\iftoggle{bbx:url}
	{\usebibmacro{url+urldate}}
	{}%
	\newunit\newblock
	\usebibmacro{addendum+pubstate}%
	\setunit{\bibpagerefpunct}\newblock
	\usebibmacro{pageref}%
	\newunit\newblock
	\usebibmacro{related}%
	\usebibmacro{finentry}}

\DeclareFieldFormat[report]{titlecase}{#1}

\DeclareFieldFormat[report]{title}{%
    \iffieldequalstr{type}{standard}{%
        \mkbibemph{#1}%
    }{%
        \mkbibquote{#1}%
    }%
}%
\DeclareFieldFormat[report]{type}{\MakeCapital{#1}}

\newbibmacro*{standardno}{%
    \printlist{institution}%
    \setunit{\addspace}%
    \iffieldundef{note}{%
        \printfield{type}%
    }{%
        \printfield{note}%
    }%
    \setunit{\addspace}
    \printfield{number}%
}

\DeclareBibliographyDriver{report}{%
  \iffieldequalstr{type}{standard}{%
          \usebibmacro{bibindex}%
          \usebibmacro{begentry}%
          \printfield{title}%
          \setunit{\labelnamepunct}\newblock
          \usebibmacro{standardno}%
          \setunit{\labelnamepunct}\newblock
          \usebibmacro{date}
  }{%
      \usebibmacro{bibindex}%
      \usebibmacro{begentry}%
      \usebibmacro{author}%
      \setunit{\labelnamepunct}\newblock
      \usebibmacro{title}%
      \newunit
      \printlist{language}%
      \newunit\newblock
      \usebibmacro{byauthor}%
      \newunit\newblock
      \usebibmacro{institution+location}%
      \newunit\newblock
      \printfield{type}%
      \setunit*{\addspace}%
      \printfield{number}%
      \newunit\newblock
      \printfield{version}%
      \newunit\newblock
      \usebibmacro{date}%
      \newunit
      \printfield{note}%
      \newunit\newblock
      \usebibmacro{chapter+pages}%
      \newunit
      \printfield{pagetotal}%
      \newunit\newblock
      \iftoggle{bbx:isbn}
        {\printfield{isrn}}
        {}%
      \newunit\newblock
      \usebibmacro{doi+eprint+url}%
      \newunit\newblock
      \usebibmacro{addendum+pubstate}%
      \setunit{\bibpagerefpunct}\newblock
      \usebibmacro{pageref}%
      \newunit\newblock
      \iftoggle{bbx:related}
        {\usebibmacro{related:init}%
         \usebibmacro{related}}
        {}%
    }
    \usebibmacro{finentry}}

\usepackage[caption=false, font=footnotesize]{subfig}
\usepackage{ifthen}
\usepackage{lipsum}
\usepackage{cleveref}

\usepackage{xstring}
\usepackage[abspath]{currfile}

\usepackage{bm} 		

\usepackage{amssymb} 	
\usepackage{bm} 		
\usepackage{booktabs} 	
\usepackage{multirow}   
\usepackage{booktabs} 	
\usepackage{colortbl}
\usepackage{array}

\usepackage{microtype}

\AtBeginDocument{\providecommand\tikzexternaldisable{\relax}} 
\AtBeginDocument{\providecommand\tikzexternalenable{\relax}} 

\newcolumntype{P}[1]{>{\centering\arraybackslash}p{#1}}
\usepackage{xstring}
\usepackage[abspath]{currfile}
\usepackage{svg}
\usepackage{adjustbox}
\input{macros}

\definecolor{mn}{RGB}{255,127,0}
\definecolor{ts}{RGB}{0,0,255}
\definecolor{mr}{RGB}{190,0,80}

\title{\LARGE \bf
A Review of Conceptualizations of Safety and Risk in Current Automated Driving Regulation$^\ast$
}

\author{Marcus Nolte$^{1}$, Leon Johann Brettin$^{1}$, Hans Steege$^{2}$, Nayel Salem$^{1}$, Marvin Loba$^{1}$,\\ Robert Graubohm$^{1}$, and Markus Maurer$^{1}$
\thanks{$^\ast$This work was supported by the German Federal Ministry for Economic Affairs and Climate Action within the project ``Automatisierter Transport zwischen
Logistikzentren auf Schnellstraßen im Level 4 (ATLAS-L4)''.}
\thanks{\hspace{-1em}$^{1}$TU Braunschweig, Institute of Control Engineering, Hans-Sommer-Str. 66, 38106 Braunschweig, Germany
        {\tt\small \{m.nolte, l.brettin, n.salem, m.loba, robert.graubohm, markus.maurer\}@tu-braunschweig.de}\vspace{0.3em}\newline%
       $^{2}$University of Stuttgart, Institute of Economics and Law 
        {\tt\small hans.steege@ivr.uni-stuttgart.de}}%
}

\begin{document}

\maketitle
\thispagestyle{empty}
\pagestyle{empty}

%
\begin{abstract}%
	``Safety'' and ``Risk'' are key concepts for the design and development of automated vehicles.
For the market introduction or large-scale field tests, both concepts are not only relevant for engineers developing the vehicles, but for all stakeholders (e.g., regulators, lawyers, or the general public) who have stakes in the technology.
In the communication between stakeholder groups, common notions of these abstract concepts are key for efficient communication and setting mutual expectations.

In the European market, automated vehicles require Europe-wide type approval or at least operating permits in the individual states.
For this, a central means of communication between regulators and engineers are regulatory documents.
Flawed terminology regarding the safety expectations for automated vehicles can unnecessarily complicate relations between regulators and manufacturers and thus hinder the introduction of the technology.
In this paper, we review relevant documents at the UN- and EU-level, for the UK, and Germany regarding their (implied) notions of safety and risk.
We contrast the regulatory notions with established and more recently developing notions of safety and risk in the field of automated driving.
Based on the analysis, we provide recommendations on how explicit definitions of safety and risk in regulatory documents can support rather than hinder the market introduction of automated vehicles.
\end{abstract}%
\copyrightnotice

\section{Introduction}
\label{sec:intro}
The market introduction of SAE-Level-3 automated driving systems in the European Union and the UK requires active regulatory acts to enable and implement type approval, commercial operation, as well as public-road testing.
The type approval system is often referred to as a way to ensure a ``safe'' market introduction of new technologies, while having a reputation of potentially slowing down innovation processes.

For automated vehicles, the need for a-priori enabling regulation creates three main challenges:
First, regulatory bodies must create laws \emph{before} a technology can be introduced in the regulated market.
This can become a competitive disadvantage when compared to target markets, such as the US, that employ a system of self-certification.
Second, regulatory bodies must establish an appropriate level of technical expertise.
This ensures that the enacted laws and regulations support the introduction of new technologies, rather than hindering them unintentionally due to lacking technical feasibility.
Third, the laws and acts must be written in a way that provides sufficient legal certainty for companies developing automated vehicles while avoiding overregulation, which could cause additional competitive disadvantages compared to other markets.

This paper will focus on the third challenge.
Concretizing regulation can be done in several ways.
One option is to provide concrete metrics and checklists that must be delivered by manufacturers to prove that the systems are ``safe enough''.
A second way of concretization is to introduce concise terminology regarding the concepts of ``safety'' and ``risk'':
One of the goals related to the type approval system, is to bind companies by regulation to prove that their systems are ``safe'' \emph{before} market introduction.

However, in everyday language, ``safety'' and ``risk'' constitute so-called ``open'' or ``empty signifiers'' \parencite{fleischer2023, buchanan2010} (cf. \cref{sec:related-work}) which means that the words do not carry precise semantic meanings in general language.
Different stakeholders in transdisciplinary (or even inter- and intradisciplinary) discussions may thus attribute their own understanding to both terms.
In consequence, stakeholders involved in discussions can seemingly agree on the importance of ``safe'' automated vehicles while having different understandings of what ``safe'' actually means.
In this regard, precise regulation requires the clear definition of such floating signifiers to ensure a common understanding of terms between regulators. 

While this understanding is key between engineers and regulators, even in engineering disciplines, the established technical notions of ``safety'' and ``risk'', derived from hazards to the health and life of humans (in short: physical harm), are currently coming under scrutiny \parencite{koopman2024}:
As automated vehicles are part of a larger sociotechnical system of systems \parencite{koopman2024,salem2024}, there are arguments that also technical notions should, e.g., include ethical or legal considerations that go beyond physical harm.

We want to raise awareness of potential issues that can be caused by differing (implied) understandings regarding the concepts of safety and risk and provide recommendations to regulators how terminology can be improved.

 This paper is organized as follows: \Cref{sec:related-work} gives an overview of current discussions regarding the perception of the concepts of ``safety'' and ``risk'' in the field of automated driving.
Based on the findings from the literature, we review current European regulation (\Cref{sec:analysis}), including the ``EU Implementing Act'' (Regulation (EU) 2022/1426 \parencite{eu1426}), the German Implementing Regulation (AFGBV \parencite{afgbv}), as well as the 2024 UK Automated Vehicles Act \parencite{ukav2024} (``UK AV Act'') regarding their understandings of the concepts of ``safety'' and ``risk''.
\Cref{sec:conclusion} provides concluding recommendations for making  regulation more concise and less prone to misunderstandings.

\section{Related Work}
\label{sec:related-work}
The need for clear terminology when it comes to communication about ``risk'' and ``safety'' is not particular to the field of automated driving.
A core concern in the fields of risk research and risk communication \parencite{renn1998} is to establish well-grounded, ideally commonly understandable \parencite{sellnow2009}, terminology between different stakeholders, e.g., the general public, regulatory bodies, policymakers, industry, or public institutions such as non-governmental organizations (NGOs) \parencite{fischhoff1984,renn1998,christensen2003}.
In this context, \citeauthor{fischhoff1984} demand ``[\ldots] an explicit and accepted definition of the term `risk'[\ldots]'' \parencite[123]{fischhoff1984}, noting that the definition of ``risk'' is ``inherently controversial'' \parencite[124]{fischhoff1984}.

\citeauthor{christensen2003} note that deviating terminology between stakeholders can derail discussions from their ``core issue(s)'' \parencite[182]{christensen2003}.
They specifically include terminology that is related to ``identifying, estimating, regulating, and communicating risk'' \parencite[182]{christensen2003}, hence including all the above-mentioned stakeholders in their argument.
The authors analyze several references from different regulatory bodies and NGOs such as the European Commission, the UN/OECD, the US-EPA, or ISO/IEC (specifically ISO/IEC Guide 51 \parencite{iso51}).
\citeauthor{christensen2003} \parencite{christensen2003} discuss, explain, and clarify applications of terms and views related to risk associated sciences to facilitate communication between stakeholder groups.
By not providing a fully consolidated terminology, they acknowledge that, while there must be a fundamental consensus among stakeholders, communication always has to be tailored to the communicating parties.
In the following sections of the paper, we will mainly consider regulators communicating to industry who is implementing automated driving technology.

The need for clear and consistent communication about risks related to autonomous systems in general is highlighted in \parencite{wmg2023}.
The report emphasizes that it is crucial to identify \emph{who} communicates \emph{how} about \emph{what} related to the communication about safety-critical autonomous systems.
As \parencite{christensen2003}, the authors explicitly stress that messages related to safety assurance and their content should be tailored to the relevant audiences.
According to \parencite{wmg2023}, this is particularly important for calibrating the expectations of different stakeholders and for raising awareness for the limitations of autonomous systems' capabilities.

Different concepts of ``risk'' and ``safety'' for automated driving systems have, e.g., been discussed in \parencite{koopman2024} or \parencite{salem2024}.
\citeauthor{koopman2024} \parencite{koopman2024} review automotive safety standards (ISO~26262:2018 \parencite{iso2018}, ISO~21448:2022 \parencite{iso21448}, ANSI/UL4600 \parencite{ul4600}) and additional resources from the German Ethics Commission \parencite{difabio2017} to the US National Highway Traffic Safety Administration for their conceptualizations of ``safety''.
They provide additional examples of ``safety problems'' related to automated driving systems which are not covered by the purely technical definitions assumed in ISO~26262 and ISO~21448.
The authors discuss that risk and safety for automated vehicles should be discussed in a more nuanced way than only considering technical definitions grounded in ``net (physical or monetary) harm''.
However, the discussion related to ``risk'' falls slightly short, not acknowledging that existing risk definitions (e.g., \parencite{fischhoff1984,renn1998}) already allow for broader discussions beyond ``net harm''.
While the newly proposed risk definition does include the influence of ``importance for stakeholders'', which allows a more interdisciplinary view on risk, the risk definition is slightly derailed by the introduction of the additional term ``loss''.
``Loss'' is defined in a slightly broader sense, compared to harm in ISO Guide~51 (see below), including ``negative societal externalities'' and ``injury and death of animals''.
While this extends technical notions of risk, it mainly shifts complexity in the terminology, addressing rather the effects of inflicted harm than the harm (violated stakeholder needs and values) itself.

In previous publications (\parencite{salem2024, nolte2024}), we discuss possible conceptualizations of risk and safety, considering a broader view of defining ``safety'' and ``risk'', particularly addressing ethical questions and stakeholder values.
While we provide a similar assessment of technical standards as \citeauthor{koopman2024}, we include general safety and risk management standards such as IEC~61508 \parencite{iec61508}, ISO~31000 \parencite{iso31000}, or ISO Guide 51 \parencite{iso51} to avoid a narrow technical focus.
Particularly, we consider ethical questions for defining ``risk'' and ``safety'' by relating harm to stakeholder values\footnote{When compared to \citeauthor{koopman2024}, harmed stakeholder values are a potential root cause for negative societal externalities that express non-acceptance of the technology.}.

Considering this body of related work, several distinctions can be made when assessing definitions for ``safety'' and ``risk'':
Notions of ``safety'' and ``risk'' can be separated by underlying ``risk sources''\footnote{E.g., E/E-failures as per ISO~26262 or functional insufficiencies as per ISO~21448.} (as per \parencite{christensen2003}; we'll argue in \cref{sec:analysis}, why we prefer the term ``hazard sources'') or the considered types of ``harm''\footnote{Such as physical or monetary harm.}
In the following, we will give definitions and perspectives on ``safety'' and ``risk'' that will be used for assessing the terminology provided in the regulatory documents.

\section{Safety \& Risk -- Terminology}
\label{sec:terminology}
For the most part, discussing ``safety'' and ``risk'' independently is impossible, as ``safety'' is typically defined with the help of ``risk''.
Safety can hence be understood as the ``freedom from (unacceptable/not tolerable/unreasonable) risk''.
Note that depending on the prefix (``unreasonable'' as in ISO~26262 \parencite[p.~21, def.~3.132]{iso2018}, ``unacceptable'' as in IEC~61508 \parencite{iec61508}), ``safety'' can take an absolute (as in ``no risk is tolerable'') or a relative (as in ``a sufficiently low risk is tolerable'') notion.

The most general ``risk'' definition is the ``effect of uncertainty on objectives'' \parencite{iso31000}.
In technical domains, this uncertainty is often reflected in terms of \emph{probability}.
The effect on objectives is considered as \emph{harm} with a certain \emph{severity}.
This yields a technical definition of ``risk'' as the ``combination of the probability of occurrence of harm [\ldots] and the severity of that harm'' \parencite[p.~2, def.~3.9]{iso51}, with ``harm'' being the ``injury or damage to the health of people, or damage to property or the environment'' \parencite[p.~1, def.~3.1]{iso51}.\footnote{Note that this still is the more materialistic interpretation of ``harm'' as criticized in \parencite{koopman2024}.}
As noted in \parencite{salem2024, nolte2024}, this technical risk definition already allows additional dimensions of ``risk'' when broadening the concept of ``harm'' to, e.g., include violations of stakeholder values such as mobility or legality.

Regarding the understanding of ``safety'' in the context of technical systems, the ``freedom from untolerable risk'' is a generally accepted notion among engineers.
Safety-related technical standards refer to a probability of harm, which suggests that, in engeinnering disciplines, it is clear that complex technical systems will never be infallible.
In addition, the general public is indeed willing to accept certain risks, as long as they see sufficient benefit from this acceptance \parencite{grunwald2016} -- How much risk is ``tolerable'', is then subject to the definition of (``risk'') acceptance criteria.
However, this willingness to accept risk can also be dependent on how professional actors communicate about the risks involved in a technology.

In this respect, other disciplines and the general public may tend towards more absolute notions of ``safety'':
The Cambridge Dictionary defines safety as ``a state in which or a place where you are safe and not in danger or at risk'' \parencite{cambridge2025}.
Despite the cyclic definition of safety by ``safe'', it is clear, that ``not being at risk'' suggests an absolute understanding regarding the absence of risk.
Ballantine's Law Dictionary defines ``safe'' as ``out of harm's way'', pointing to concepts such as ``reasonably safe'' or ``adequately safe'' for more relaxed definitions.
With that, the basic definition proposes no chance of occurring harm and no possible relaxation of the concept of ``safety'' that can be related to risk.
In German jurisprudence, \parencite{peine1999} defines that a risk is given by a ``theoretically possible, but practically unlikely'' harm, also pointing to a rather absolute notion of safety.

Regarding automated driving systems, these nuances are important:
In addition to the risks caused by failures in complex technical systems, automated vehicles are operating in a practically open world. \parencite{nolte2024, koopman2024} 
Developers thus cannot foresee every possible scenario or required system reaction at design time.
Implementing overly cautious vehicle behavior as a safety measure would sacrifice mobility \parencite{graubohm2023} and by that reduce to utility of the technology.
If automated vehicles shall provide mobility to their users, developers need to find a balance between a system design that is sufficiently safe and provides mobility at the same time.
This means, however, that the risk incurred by automated driving systems can only be mitigated (to a tolerable degree), but never be fully eliminated.

Regulation that is written with an absolute notion of safety in mind can hence a) cause legal insecurity for the developers of automated vehicles who must strike that balance between safety and mobility or b) in the worst case hinder the introduction of the technology completely.

\section{Analysis}
\label{sec:analysis}
In the following sections, we will analyze European type approval regulation\footnote{Strictly speaking, the German enabling act (AFGBV) does not regulate type-approval, but how test \& operating permits are issued for SAE-Level-4 systems. Type-approval regulation for SAE-Level-3 systems follows UN Regulation No. 157 (UN-ECE-ALKS) \parencite{un157}.} regarding the underlying notions of ``safety'' and ``risk''.
We will classify these notions according to their absolute or relative character, underlying risk sources, or underlying concepts of harm.

\subsection{Classification of Safety Notions}
\label{sec:safety-notions}
We will refer to \emph{absolute} notions of safety as conceptualizations that assume the complete absence of any kind of risk.
Opposed to this, \emph{relative} notions of safety are based on a conceptualization that specifically includes risk acceptance criteria, e.g., in terms of ``tolerable'' risk or ``sufficient'' safety.

\begin{figure*}
    \includeimg[width=.9\textwidth]{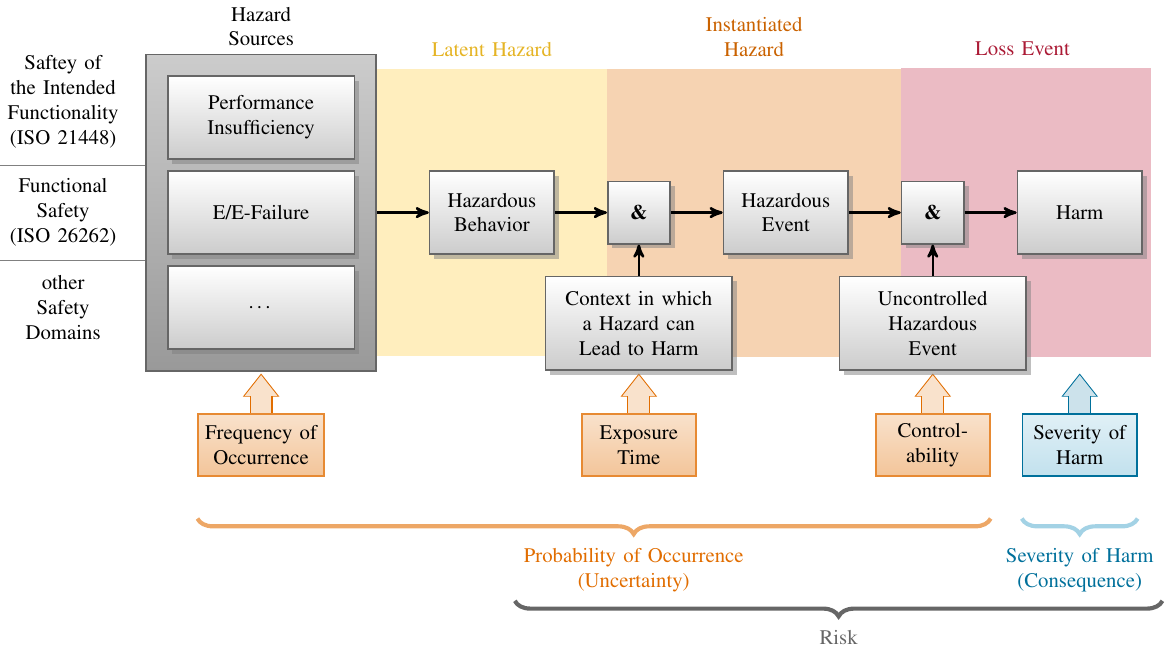}
    \caption{Graphical summary of a taxonomy of risk related to automated vehicles, extended based on ISO 21448 (\parencite{iso21448}) and \parencite{favaro2023}. Top: Causal chain from hazard sources to actual harm; bottom: summary of the individual elements' contributions to a resulting risk. Graphic translated from \parencite{nolte2024} \label{fig:hazard-sources}}
\end{figure*}

For classifying notions of safety by their underlying risk (or rather ``hazard'') sources, and different concepts of harm, \Cref{fig:hazard-sources} provides an overview of our reasoning, which is closely in line with the argumentation provided by Waymo in \parencite{favaro2023}.
We prefer ``hazard sources'' over ``risk sources'', as a risk must always be related to a \emph{cause} or \emph{source of harm} (i.e., a hazard \parencite[p.~1, def. 3.2]{iso51}).
Without a concrete (scenario) context that the system is operating in, a hazard is \emph{latent}: E.g., when operating in public traffic, there is a fundamental possibility that a \emph{collision with a pedestrian} leads to (physical) harm for that pedestrian. 
However, only if an automated vehicle shows (potentially) hazardous behavior (e.g., not decelerating properly) \emph{and} is located near a pedestrian (context), the hazard is instantiated and leads to a hazardous event.
If the hazardous event cannot be mitigated or controlled, we see a loss event in which the pedestrian's health is harmed.
Note that this hypothetical chain of events is summarized in the definition of risk:
The probability of occurrence of harm is determined by a) the frequency with which hazard sources manifest, b) the time for which the system operates in a context that exposes the possibility of harm, and c) by the probability with which a hazardous event can be controlled.
A risk can then be determined as a function of the probability of harm and the severity of the harm potentially inflicted on the pedestrian.

In the following, we will apply this general model to introduce different types of hazard sources and also different types of harm.
\cref{fig:hazard-sources} shows two distinct hazard sources, i.e., functional insufficiencies and E/E-failures that can lead to hazardous behavior.
ISO~21488 \parencite{iso21448} defines functional insufficiencies as insufficiencies that stem from an incomplete or faulty system specification (specification insufficiencies).
In addition, the standard considers insufficiencies that stem from insufficient technical capability to operate inside the targeted Operational Design Domain (performance insufficiencies).
Functional insufficiencies are related to the ``Safety of the Intended Functionality (SOTIF)'' (according to ISO~21448), ``Behavioral Safety'' (according to Waymo \parencite{waymo2018}), or ``Operational Safety'' (according to UN Regulation No. 157 \parencite{un157}).
E/E-Failures are related to classic functional safety and are covered exhaustively by ISO~26262 \parencite{iso2018}.
Additional hazard sources can, e.g., be related to malicious security attacks (ISO~21434), or even to mechanical failures that should be covered (in the US) in the Federal Motor Vehicle Safety Standards (FMVSS).

For the classification of notions of safety by the related harm, in \parencite{salem2024, nolte2024}, we take a different approach compared to \parencite{koopman2024}:
We extend the concept of harm to the violation of stakeholder \emph{values}, where values are considered to be a ``standard of varying importance among other such standards that, when combined, form a value pattern that reduces complexity for stakeholders [\ldots] [and] determines situational actions [\ldots].'' \parencite{albert2008}
In this sense, values are profound, personal determinants for individual or collective behavior.
The notion of values being organized in a weighted value pattern shows that values can be ranked according to importance.
For automated vehicles, \emph{physical wellbeing} and \emph{mobility} can, e.g., be considered values which need to be balanced to achieve societal acceptance, in line with the discussion of required tradeoffs in \cref{sec:terminology}.
For the analysis of the following regulatory frameworks, we will evaluate if the given safety or risk notions allow tradeoffs regarding underlying stakeholder values. 

\subsection{UN Regulation No. 157 \& European Implementing Regulation (EU) 2022/1426}
\label{sec:enabling-act}
UN Regulation No. 157 \parencite{un157} and the European Implementing Regulation 2022/1426 \parencite{eu1426} provide type approval regulation for automated vehicles equipped with SAE-Level-3 (UN Reg. 157) and Level 4 (EU 2022/1426) systems on an international (UN Reg. 157) and European (EU 2022/1426) level.

Generally, EU type approval considers UN ECE regulations mandatory for its member states ((EU) 2018/858, \parencite{eu858}), while the EU largely forgoes implementing EU-specific type approval rules, it maintains the right to alter or to amend UN ECE regulation \parencite{eu858}.

In this respect, the terminology and conceptualizations in the EU Implementing Act closely follow those in UN Reg. No. 157.
The EU Implementing Act gives a clear reference to UN Reg. No. 157 \parencite[][Preamble,  Paragraph 1]{eu1426}.
Hence, the documents can be assessed in parallel.
Differences will be pointed out as necessary.

Both acts are written in rather technical language, including the formulation of technical requirements (e.g., regarding deceleration values or speeds in certain scenarios).
While providing exhaustive definitions and terminology, neither of both documents provide an actual definition of risk or safety.
The definition of ``unreasonable'' risk in both documents does not define risk, but only what is considered \emph{unreasonable}. It states that the ``overall level of risk for [the driver, (only in UN Reg. 157)] vehicle occupants and other road users which is increased compared to a competently and carefully driven manual vehicle.''
The pertaining notions of safety and risk can hence only be derived from the context in which they are used.

\subsubsection{Absolute vs. Relative Notions of Safety}
In line with the technical detail provided in the acts, both clearly imply a \emph{relative} notion of safety and refer to the absence of \emph{unreasonable} risk throughout, which is typical for technical safety definitions.

Both acts require sufficient proof and documentation that the to-be-approved automated driving systems are ``free of unreasonable safety risks to vehicle occupants and other road users'' for type approval.\footnote{As it targets SAE-Level-3 systems, UN Reg. 157 also refers to the driver, where applicable.}
In this respect, both acts demand that the manufacturers perform verification and validation activities for performance requirements that include ``[\ldots] the conclusion that the system is designed in such a way that it is free from unreasonable risks [\ldots]''.
Additionally, \emph{risk minimization} is a recurring theme when it comes to the definition of Minimum Risk Maneuvers (MRM).

Finally, supporting the relative notions of safety and risk, UN Reg. 157 introduces the concept of ``reasonable foreseeable and preventable'' \parencite[Article 1, Clause 5.1.1.]{un157} collisions, which implies that a residual risk will remain with the introduction of automated vehicles.
\parencite[][Appendix 3, Clause 3.1.]{un157} explicitly states that only \emph{some} scenarios that are unpreventable for a competent human driver can actually be prevented by an automated driving system.
While this concept is not applied throughout the EU Implementing Act, both documents explicitly refer to \emph{residual} risks that are related to the operation of automated driving systems (\parencite[][Annex I, Clause 1]{un157}, \parencite[][Annex II, Clause 7.1.1.]{eu1426}).

\subsubsection{Hazard Sources}
Hazard sources that are explicitly differentiated in UN Reg. 157 and (EU) 2022/1426 are E/E-failures that are in scope of functional safety (ISO~26262) and functional insufficiencies that are in scope of behavioral (or ``operational'') safety (ISO~21448).
Both documents consistently differentiate both sources when formulating requirements.

While the acts share a common definition of ``operational'' safety (\parencite[][Article 2, def. 30.]{eu1426}, \parencite[][Annex 4, def. 2.15.]{un157}), the definitions for functional safety differ.
\parencite{un157} defines functional safety as the ``absence of unreasonable risk under the occurrence of hazards caused by a malfunctioning behaviour of electric/electronic systems [\ldots]'', \parencite{eu1426} drops the specification of ``electric/electronic systems'' from the definition.
When taken at face value, this definition would mean that functional safety included all possible hazard sources, regardless of their origin, which is a deviation from the otherwise precise usage of safety-related terminology.

\subsubsection{Harm Types}
As the acts lack explicit definitions of safety and risk, there is no consistent and explicit notion of different harm types that could be differentiated.

\parencite{un157} gives little hints regarding different considered harm types.
``The absence of unreasonable risk'' in terms of human driving performance could hence be related to any chosen performance metric that allows a comparison with a competent careful human driver including, e.g., accident statistics, statistics about rule violations, or changes in traffic flow.

In \parencite{eu1426}, ``safety'' is, implicitly, attributed to the absence of unreasonable risk to life and limb of humans.
This is supported by the performance requirements that are formulated:
\parencite[][Annex II, Clause 1.1.2. (d)]{eu1426} demands that an automated driving system can adapt the vehicle behavior in a way that it minimizes risk and prioritizes the protection of human life.

Both acts demand the adherence to traffic rules (\parencite[][Annex 2, Clause 1.3.]{eu1426}, \parencite[][Clause 5.1.2.]{un157}).
\parencite[][Annex II, Clause 1.1.2. (c)]{eu1426} also demands that an automated driving system shall adapt its behavior to surrounding traffic conditions, such as the current traffic flow.
With the relative notion of risk in both acts, the unspecific clear statement that there may be unpreventable accidents \parencite{un157}, and a demand of prioritization of human life in \parencite{eu1426}, both acts could be interpreted to allow developers to make tradeoffs as discussed in \cref{sec:terminology}.

\subsubsection{Conclusion}
To summarize, the UN Reg. 157 and the (EU) 2022/1426 both clearly support the technical notion of safety as the absence of unreasonable risk.
The notion is used consistently throughout both documents, providing a sufficiently clear terminology for the developers of automated vehicles.
Uncertainty remains when it comes to considered harm types: Both acts do not explicitly allow for broader notions of safety, in the sense of \parencite{koopman2024} or \parencite{salem2024}.
Finally, a minor weak spot can be seen in the definition of risk acceptance criteria: Both acts take the human driving performance as a baseline.
While (EU) 2022/1426 specifies that these criteria are specific to the systems' Operational Design Domain \parencite[][Annex II, Clause 7.1.1.]{eu1426}, the reference to the concrete Operational Design Domain is missing in UN Reg. 157.
Without a clearly defined notion of safety, however, it remains unclear, how aspects beyond net accident statistics (which are given as an example in \parencite[][Annex II, Clause 7.1.1.]{eu1426}), can be addressed practically, as demanded by \parencite{koopman2024}.

\subsection{German Regulation (StVG \& AFGBV)}
\label{sec:afgbv}
The German L3 (Automated Driving Act) and L4 (Act on Autonomous Driving) Acts from 2017 and 2021,\footnote{Formally, these are amendments to the German Road Traffic Act (StVG): 06/21/2017, BGBl. I p. 1648, 07/12/2021 BGBl. I p. 3108.} respectively, provide enabling regulation for the operation of SAE-Level-3 and 4 vehicles on German roads.
The German Implementing Regulation (\parencite{afgbv}, AFGBV) defines how this enabling regulation is to be implemented for granting testing permits for SAE-Level-3 and -4 and driving permits for SAE-Level-3 and -4 automated driving systems.\footnote{Note that these permits do not grant EU-wide type approval, but serve as a special solution for German roads only. At the same time, the AFGBV has the same scope as (EU) 2022/1426.}
With all three acts, Germany was the first country to regulate the approval of automated vehicles for a domestic market.
All acts are subject to (repeated) evaluation until the year 2030 regarding their impact on the development of automated driving technology.
An assessment of the German AFGBV and comparisons to (EU) 2022/1426 have been given in \cite{steininger2022} in German.

Just as for UN Reg. 157 and (EU) 2022/1426, neither the StVG nor the AFGBV provide a clear definition of ``safety'' or ``risk'' -- even though the "safety" of the road traffic is one major goal of the StVG and StVO.
Again, different implicit notions of both concepts can only be interpreted from the context of existing wording.
An additional complication that is related to the German language is that ``safety'' and ``security'' can both be addressed as ``Sicherheit'', adding another potential source of unclarity.
Literal Quotations in this section are our translations from the German act.

\subsubsection{Absolute vs. Relative Notions of Safety}
For assessing absolute vs. relative notions of safety in German regulation, it should be mentioned that the main goal of the German StVO is to ensure the ``safety and ease of traffic flow'' -- an already diametral goal that requires human drivers to make tradeoffs.\footnote{For human drivers, this also creates legal uncertainty which can sometimes only be settled in a-posteriori court cases.}
While UN and EU regulation clearly shows a relative notion of safety\footnote{And even the StVG contains sections that use wording such as ``best possible safety for vehicle occupants'' (§1d (4) StVG) and acknowledges that there are unavoidable hazards to human life (§1e (2) No. 2c)).}, the German AFGBV contains ambiguous statements in this respect:
Several paragraphs contain a demand for a hazard free operation of automated vehicles.
§4 (1) No. 4 AFGBV, e.g., states that ``the operation of vehicles with autonomous driving functions must neither negatively impact road traffic safety or traffic flow, nor endanger the life and limb of persons.''
Additionally, §6 (1) AFGBV states that the permits for testing and operation have to be revoked, if it becomes apparent that a ``negative impact on road traffic safety or traffic flow, or hazards to the life and limb of persons cannot be ruled out''.
The same wording is used for the approval of operational design domains regulated in §10 (1) No. 1.
A particularly misleading statement is made regarding the requirements for technical supervision instances which are regulated in §14 (3) AFGBV which states that an automated vehicle has to be  ``immediately removed from the public traffic space if a risk minimal state leads to hazards to road traffic safety or traffic flow''.
Considering the argumentation in \cref{sec:terminology}, that residual risks related to the operation of automated driving systems are inevitable, these are strong statements which, if taken at face value, technically prohibit the operation of automated vehicles.
It suggests an \emph{absolute} notion of safety that requires the complete absence of risk.  
The last statement above is particularly contradictory in itself, considering that a risk \emph{minimal} state always implies a residual risk.

In addition to these absolute safety notions, there are passages which suggest a relative notion of safety:
The approval for Operational Design Domains is coupled to the proof that the operation of an automated vehicle ``neither negatively impacts road traffic safety or traffic flow, nor significantly endangers the life and limb of persons beyond the general risk of an impact that is typical of local road traffic'' (§9 (2) No. 3 AFGBV).
The addition of a relative risk measure ``beyond the general risk of an impact'' provides a relaxation (cf. also \cite{steininger2022}, who criticizes the aforementioned absolute safety notion) that also yields an implicit acceptance criterion (\emph{statistically as good as} human drivers) similar to the requirements stated in UN Reg. 157 and (EU) 2022/1426.

Additional hints for a relative notion of safety can be found in Annex 1, Part 1, No. 1.1 and Annex 1, Part 2, No. 10.
Part 1, No 1.1 specifies collision-avoidance requirements and acknowledges that not all collisions can be avoided.\footnote{The same is true for Part 2, No. 10, Clause 10.2.5.}
Part 2, No. 10 specifies requirements for test cases.
It demands that test cases are suitable to provide evidence that the ``safety of a vehicle with an autonomous driving function is increased compared to the safety of human-driven vehicles''.
This does not only acknowledge residual risks, but also yields an acceptance criterion (\emph{better} than human drivers) that is different from the implied acceptance criterion given in §9 (2) No. 3 AFGBV.

\subsubsection{Hazard Sources}
Regarding hazard sources, Annex 1 and 3 AFGBV explicitly refer to ISO~26262 and ISO~21448 (or rather its predecessor ISO/PAS~21448:2019).
However, regarding the discussion of actual hazard sources, the context in which both standards are mentioned is partially unclear:
Annex 1, Clause 1.3 discusses requirements for path and speed planning.
Clause 1.3 d) demands that in intersections, a Time to Collision (TTC) greater than 3 seconds must be guaranteed.
If manufacturers deviate from this, it is demanded that ``state-of-the-art, systematic safety evaluations'' are performed.
Fulfillment of the state of the art is assumed if ``the guidelines of ISO~26262:2018-12 Road Vehicles -- Functional Safety are fulfilled''.
Technically, ISO~26262 is not suitable to define the state of the art in this context, as the requirements discussed fall in the scope of operational (or behavioral) safety (ISO~21448).
A hazard source ``violated minimal time to collision'' is clearly a functional insufficiency, not an E/E-failure.

Similar unclarity presents itself in Annex 3, Clause 1 AFGBV: 
Clause 1 specifies the contents of the ``functional specification''.
The ``specification of the functionality'' is an artifact which is demanded in ISO~21448:2022 (Clause 5.3) \parencite{iso21448}.
However, Annex 3, Clause 1 AFGBV states that the ``functional specification'' is considered to comply to the state of the art, if the ``functional specification'' adheres to ISO~26262-3:2018 (Concept Phase).
Again, this assumes SOTIF-related contents as part of ISO~26262, which introduces the ``Item Definition'' as an artifact, which is significantly different from the ``specification of the functionality'' which is demanded by ISO~21448.
Finally, Annex 3, Clause 3 AFGBV demands a ``documentation of the safety concept'' which ``allows a functional safety assessment''.
A safety concept that is related to operational / behavioral safety is not demanded.
Technically, the unclarity with respect to the addressed harm types lead to the fact that the requirements provided by the AFGBV do not comply with the state of the art in the field, providing questionable regulation.

\subsubsection{Harm Types}
Just like UN Reg. 157 and (EU) 2022/1426, the German StVG and AFGBV do not explicitly differentiate concrete harm types for their notions of safety.
However, the AFGBV mentions three main concerns for the operation of automated vehicles which are \emph{traffic flow} (e.g., §4 (1) No. 4 AFGBV), compliance to \emph{traffic law} (e.g., §1e (2) No. 2 StVG), and the \emph{life and limb of humans} (e.g., §4 (1) No. 4 AFGBV).

Again, there is some ambiguity in the chosen wording:
The conflict between traffic flow and safety has already been argued in \cref{sec:terminology}.
The wording given in §4 (1) No. 4 and §6 (1) AFGBV  demand to ensure (absolute) safety \emph{and} traffic flow at the same time, which is impossible (cf. \cref{sec:terminology}) from an engineering perspective.
§1e (2) No. 2 StVG defines that ``vehicles with an autonomous driving function must [\ldots] be capable to comply to [\ldots] traffic rules in a self-contained manner''.
Taken at face value, this wording implies that an automated driving system could lose its testing or operating permit as soon as it violates a traffic rule.
A way out could be provided by §1 of the German Traffic Act (StVO) which demands careful and considerate behavior of all traffic participants and by that allows judgement calls for human drivers.
However, if §1 is applicable in certain situations is often settled in court cases. 
For developers, the application of §1 StVO during system design hence remains a legal risk.

While there are rather absolute statements as mentioned above, sections of the AFGBV and StVG can be interpreted to allow tradeoffs:
§1e (2) No. 2 b) demands that a system,  ``in case of an inevitable, alternative harm to legal objectives, considers the significance of the legal objectives, where the protection of human life has highest priority''.
This exact wording \emph{could} provide some slack for the absolute demands in other parts of the acts, enabling tradeoffs between (tolerable) risk and mobility as discussed in \cref{sec:terminology}.
However, it remains unclear if this interpretation is legally possible.

\subsubsection{Conclusion}
Compared to UN Reg. 157 and (EU) 2022/1426, the German StVG and AFGBV introduce openly inconsistent notions of safety and risk which are partially directly contradictory:
The wording partially implies absolute and relative notions of safety and risk at the same time.
The implied validation targets (``better'' or ``as good as'' human drivers) are equally contradictory. 
The partially implied absolute notions of safety, when taken at face value, prohibit engineers from making the tradeoffs required to develop a system that is safe and provides customer benefit at the same time. 
In consequence, the wording in the acts is prone to introducing legal uncertainty.
This uncertainty creates additional clarification need and effort for manufacturers and engineers who design and develop SAE-Level-3 and -4 automated driving systems. The use of undefined legal terms not only makes it more difficult for engineers to comply with the law, but also complicates the interpretation of the law and leads to legal uncertainty.

\subsection{UK Automated Vehicles Act 2024 (2024 c. 10)}
The UK has issued a national enabling act for regulating the approval of automated vehicles on the roads in the UK.
To the best of our knowledge, concrete implementing regulation has not been issued yet.
Regarding terminology, the act begins with a dedicated terminology section to clarify the terms used in the act \parencite[Part 1, Chapter 1, Section 1]{ukav2024}.
In that regard, the act defines a vehicle to drive ```autonomously' if --- (a)
it is being controlled not by an individual but by equipment of the vehicle, and (b) neither the vehicle nor its surroundings are being monitored by an individual with a view to immediate intervention in the driving of the vehicle.''
The act hence covers SAE-Level-3 to SAE-Level-5 automated driving systems.

\subsubsection{Absolute vs. Relative Notions of Safety}
While not providing an explicit definition of safety and risk, the UK Automated Vehicles Act (``UK AV Act'') \parencite{ukav2024} explicitly refers to a relative notion of safety.
Part~1, Chapter~1, Section~1, Clause (7)~(a) defines that an automated vehicle travels ```safely' if it travels to an acceptably safe standard''.
This clarifies that absolute safety is not achievable and that acceptance criteria to prove the acceptability of residual risk are required, even though a concrete safety definition is not given.
The act explicitly tasks the UK Secretary of State\footnote{Which means, that concrete implementation regulation needs to be enacted.} to install safety principles to determine the ``acceptably safe standard'' in Part~1, Chapter~1, Section~1, Clause (7)~(a).
In this respect, the act also provides one general validation target as it demands that the safety principles must ensure that ``authorized automated vehicles will achieve a level of safety equivalent to, or higher than, that of careful and competent human drivers''.
Hence, the top-level validation risk acceptance criterion assumed for UK regulation is ``\emph{at least as good} as human drivers''.

\subsubsection{Hazard Sources}
The UK AV Act contains no statements that could be directly related to different hazard sources.
Note that, in contrast to the rest of the analyzed documents, the UK AV Act is enabling rather than implementing regulation.
It is hence comparable to the German StVG, which does not refer to concrete hazard sources as well.

\subsubsection{Types of Harm}
Even though providing a clear relative safety notion, the missing definition of risk also implies a lack of explicitly differentiable types of harm.
Implicitly, three different types of harm can be derived from the wording in the act.
This includes the harm to life and limb of humans\footnote{Part~1, Chapter~3, Section~25 defines ``aggravated offence where death or serious injury occurs'' \parencite{ukav2024}.}, the violation of traffic rules\footnote{Part~1, Chapter~1, Clause~(7)~(b) defines that an automated vehicle travels ```legally' if it travels with an acceptably low risk of committing a traffic infraction''}, and the cause of inconvenience to the public \parencite[Part~1, Chapter~1, Section~58, Clause (2)~(d)]{ukav2024}.

The act connects all the aforementioned types of harm to ``risk'' or ``acceptable safety''.
While the act generally defines criminal offenses for providing ``false or misleading information about safety'', it also acknowledges possible defenses if it can be proven that ``reasonable precautions'' were taken and that ``due diligence'' was exercised to ``avoid the commission of the offence''.
This statement could enable tradeoffs within the scope of ``reasonable risk'' to the life and limb of humans, the violation of traffic rules, or to the cause of inconvenience to the public, as we argued in \cref{sec:terminology}.

\subsubsection{Conclusion}
From the set of reviewed documents, the current UK AV Act is the one with the most obvious relative notions of safety and risk and the one that seems to provide a legal framework for permitting tradeoffs.
In our review, we did not spot major inconsistency beyond a missing definitions of safety and risk\footnote{Note that with the Office for Product Safety and Standards (OPSS), there is a British government agency that maintains an exhaustive and widely focussed ``Risk Lexicon'' that provides suitable risk definitions. For us, it remains unclear, to what extent this terminology is assumed general knowledge in British legislation.}.
The general, relative notion of safety and the related alleged ability for designers to argue well-founded development tradeoffs within the legal framework could prove beneficial for the actual implementation of automated driving systems.
While the act thus appears as a solid foundation for the market introduction of automated vehicles, without accompanying implementing regulation, it is too early to draw definite conclusions.
\section{Conclusion \& Recommendations}
\label{sec:conclusion}
This paper reviewed UN, EU, German, and UK regulation for the type approval of SAE-Level-$\geq$3 automated driving systems regarding the underlying notions of safety and risk.
Our analysis has shown that the considered acts vary significantly in the consistency of the used concepts.
The UK AV Act appeared as the regulation with the most consistent, yet generic, conceptualization of safety and risk.
The UN and EU regulations adopt a very technical notion of safety and risk, based on the absence of unreasonable risk, while mostly focusing on harm to the life and limb of humans.
Compared to the other acts, the German AFGBV provides ambiguous notions of risk and safety.
These notions range from the assumption of a \emph{complete} absence of risk, to more technical notions considering the absence of \emph{unreasonable} risk.

In summary, UN and EU regulations as well as the UK AV Act provide frameworks that permit traditional technical interpretations of risk management.
Moreover, the UK AV Act could provide a legal framework that allows adopting more holistic perspectives on safety, as discussed by \parencite{salem2024} or \parencite{koopman2024}.

The German StVG and AFGBV introduce significant legal uncertainty for designers and developers of automated vehicles due to ambiguous or even contradictory wording and requirements.
This creates additional effort for engineers to validate assumptions regarding the requirements with the regulators.
However, as the regulatory statutes stand as they are, even such clarification can entail the need for court cases, before legal uncertainty is settled.
This can unnecessarily impede the market introduction of automated vehicles, where clear regulation could have helped from the beginning.

Finally, specifically as the German acts are subject to regular evaluation and adaption, we recommend regulators
\begin{enumerate}
    \item[a)] to issue clarifying regulation to fix contradictions and ambiguity, particularly regarding open signifiers such as ``safety'' and ``risk'',
    \item[b)] to consider how a broader notion of risk connected to different harm types can force developers to consider negative impacts such as ``inconvenience for the public'' (see \parencite{koopman2024} for more examples), and
    \item[c)] to find wordings that encourage designers and developers to diligently document the rationale for development decisions that are involved in the required tradeoffs between stakeholder needs and values. 
\end{enumerate}

The UK AV Act could serve as an interesting example in this regard, despite only providing enabling regulation and despite the lack of explicit definitions for risk and safety.

%
\renewcommand*{\bibfont}{\footnotesize} 
\printbibliography 
\end{document}